\documentclass{easychair}

\usepackage[T1]{fontenc}
\def\doi#1{\href{https://doi.org/\detokenize{#1}}{\url{https://doi.org/\detokenize{#1}}}}
\usepackage{graphicx}

\usepackage{bm}

\usepackage{booktabs}

\usepackage{xcolor}
\usepackage{hyperref}

\usepackage{listings}
\usepackage{inconsolata} 

\lstdefinestyle{mystyle}{
    language=Python,
    frame=single,
    numbers=right,
    numbersep=5pt,
    captionpos=b,
    showstringspaces=false,
    basicstyle=\ttfamily\footnotesize,  
}

\usepackage{booktabs}

\usepackage{amsfonts}

\usepackage[numbers,square]{natbib}

\usepackage[detect-weight=true]{siunitx}

\DeclareSIUnit{\nothing}{\relax}
\DeclareSIUnit{\instr}{i}

\usepackage{xspace}

\newcommand\solver[1]{\textsc{#1}\xspace}
\newcommand\vampire{\solver{Vampire}}

\newcommand\probtptp[1]{\texttt{#1}}

\newcommand\dataset[1]{\textit{#1}\xspace}
\newcommand\tptp{\dataset{TPTP}}
\newcommand\mizar{\dataset{Mizar40}}
\newcommand\isabelle{\dataset{Isabelle}}
\newcommand\coqhammer{\dataset{CoqHammer}}

\renewcommand{\ttdefault}{cmtt} 

\newcommand\supershorten[1]{}

\makeatletter
\renewcommand\section{\@startsection{section}{1}{\z@}%
                       {-12\p@ \@plus -4\p@ \@minus -4\p@}%
                       {8\p@ \@plus 4\p@ \@minus 4\p@}%
                       {\normalfont\large\bfseries\boldmath
                        \rightskip=\z@ \@plus 8em\pretolerance=10000 }}
\makeatother

\begin{document}
\title{
Teaching Vampire New Tricks: An Experimental Study of Neural Clause Selection
}
\titlerunning{Teaching Vampire New Tricks}
\author{Karel Chvalovský\inst{1} 
\and
Martin Suda\inst{1} 
\and
Josef Urban\inst{2} 
}
\authorrunning{K. Chvalovský, M. Suda, and J. Urban}
%
\institute{Czech Technical University in Prague, Czech Republic \\
    \email{\{karel.chvalovsky,martin.suda\}@cvut.cz}
\and
AI4REASON and University of Gothenburg \\ \email{josef.urban@gmail.com}
}

\maketitle              

\begin{abstract}
A neural clause-selection guidance approach in the \vampire{} theorem prover was recently shown
to substantially improve the success rate of the prover's default strategy on the TPTP benchmark.
We experimentally study the impact of the approach across several ITP-derived benchmark sets
and its interaction with theorem proving strategies.

We find that while the neural guidance consistently improves performance within individual benchmark domains,
cross-bench\-mark application of guiding models underperforms the plain default strategy.
This can be remedied by training a single model on all datasets at once. Such a model,
although more expensive to obtain, helps  \vampire{} almost catch up in performance across all datasets.
The picture when considering combined strategies is less clear-cut, indicating 
persisting value of neural guidance but under diminishing returns.
\end{abstract}



\renewcommand{\ttdefault}{zi4} 

\section{Introduction}

Neural guidance has recently become a competitive technique for improving
saturation-based automated theorem provers \cite{DBLP:conf/cade/Suda25,DBLP:conf/lpar/ChvalovskyKPU23,DBLP:conf/itp/JakubuvCGKOP00U23}. In particular, a neural
clause-{se\-lec\-tion} guidance integrated into the \vampire{} \cite{DBLP:conf/cav/BartekBCHHHKRRRSSV25} prover
has been shown to substantially improve the performance of its default
strategy on the TPTP \cite{Sut17} benchmark library.
The approach \cite{DBLP:conf/cade/Suda25}, inspired by reinforcement-learning (RL), replaces handcrafted clause-ranking heuristics with a
graph-based clause-scoring neural model iteratively trained on proof traces, yielding significant
improvements in terms of the number of problems solved under strict resource limits.
These results demonstrate that learned clause selection can
outperform carefully engineered heuristics in a fixed setting. However, several
important research questions remain open:
\begin{enumerate}
\item \textbf{Broader effectiveness:} Neural clause-selection guidance has so far been shown to boost
a single strategy on the TPTP benchmark. Modern automated theorem provers are
increasingly deployed as \emph{hammers} \cite{DBLP:journals/jfrea/BlanchetteKPU16} for interactive theorem provers
(ITPs), where problem distributions differ substantially from classical
ATP benchmarks. Does the observed improvement carry over to these structurally different
problem distributions? Confirming (or refuting) this on benchmarks exported from
ITPs is a result in its own right and a necessary first step to also answering the next question.
\item \textbf{Transfer:} Given multiple benchmark families with different characteristics,
how well does neural clause-selection guidance generalize beyond
the dataset on which it is trained? Understanding whether models
trained on one benchmark family transfer to others is important
for assessing the practical reach of the approach.
\item \textbf{Portfolios:} Prior work focused on boosting a single prover configuration.
In practice, however, state-of-the-art ATP performance is achieved via
portfolios of complementary strategies and carefully constructed
schedules. This raises the question of how neural clause selection
interacts with strategy diversity: does it merely compensate for
weaknesses of a particular configuration, or does it systematically
enhance optimized strategies and improve portfolio performance?
\end{enumerate}

In this work, we conduct a comprehensive experimental study of neural
clause-selection guidance across four benchmark families: TPTP,
Mizar40 \cite{DBLP:journals/jar/KaliszykU15a}, Isabelle/Sledgehammer \cite{DBLP:conf/cade/BohmeN10}
exports \cite{DBLP:conf/itp/GoertzelJKOPU22}, and CoqHammer \cite{DBLP:journals/jar/CzajkaK18}. Building on the
RL-inspired framework of~\cite{DBLP:conf/cade/Suda25}, we
analyze its per-dataset performance gains and cross-dataset transfer.
Moreover, we study its interaction with locally optimized strategies and its impact on large-scale
portfolio construction via greedy cover and spider-style strategy
search \cite{DBLP:conf/ijcar/BartekCS24}. Our main results can be summarized as follows:
\begin{itemize}
\item
Neural guidance consistently improves performance within individual domains, often substantially on harder hammer-derived
benchmarks. However, models trained on one dataset typically fail to transfer effectively to others, frequently underperforming the default
unguided strategy. This can be remedied, at an increased computational cost, by training a single model on all datasets of interest.
\item
While neural boosting strongly enhances individual strategies (and a single neural model can be trained to successfully guide more than one strategy), 
the marginal benefit of guidance diminishes in larger portfolios where strategic complementarity already captures much of the feasible-to-claim part of the
search space.
\item
Inspired by Goertzel's experiment \cite{DBLP:conf/cade/Goertzel20} with E prover \cite{DBLP:conf/cade/0001CV19} and ENIGMA clause-selection guidance \cite{DBLP:conf/cade/ChvalovskyJ0U19}, we show that neural guidance in \vampire{} can compensate for the inefficiency of unordered resolution and plain paramodulation, 
lifting the performance of such a weakened strategy above that of the unguided default (which uses the superposition calculus \cite{DBLP:conf/cade/BachmairG90,DBLP:journals/logcom/BachmairG94} and a simplification ordering).
This demonstrates that learned clause selection can tame a more prolific but theoretically interesting calculus.
\item
After 20 days of spider-style strategy search, we built greedy portfolios from plain, neural, and combined strategy pools. 
The combined portfolio covers \SI{4.1}{\percent} more problems overall but, more importantly, achieves \SI{12.2}{\percent} improvement at the practically relevant 
$\sim$\SI{10}{\second} 
time limit. This is the first evaluation of neural clause selection at the portfolio scale.
\end{itemize}
These findings clarify both the strengths and limitations
of neural clause selection in modern saturation-based theorem proving.

After recalling the neural guidance architecture and how it is trained in an iterative fashion (Sect.~\ref{sec:background}), we describe our four datasets (Sect.~\ref{sec:datasets})
and introduce the details of our experimental setup (Sect.~\ref{sec:expSetup}). With the experiments themselves,
we start by measuring the per-dataset improvement by neural guidance of \vampire{}'s default strategy and
the knowledge transfer potential of the trained models (Sect.~\ref{sec:whereDefaultModelGetsBorn}).
Moving on to theorem proving strategies (Sect.~\ref{sec:strategies}), we present a simple optimization algorithm
to give us a small set of complementary strategies (\ref{subsect:hillClimbGreed})
to which we apply neural guidance both individually (\ref{subsec:fiveStratBoost}) and jointly (\ref{subsect:singleGuidingModel}).
We pause to boost the mentioned plain paramodulation strategy (\ref{subsect:smartAgain}),
and then build and compare full-fledged portfolios both plain and neurally guided (\ref{subsect:spiderStyleStrategySearch} and \ref{subsect:greedyPortfolio}).
Last, we combine the dataset view with the strategy view (Sect.~\ref{sec:stratsAndBoosting}).
We then recall related work (Sect.~\ref{sec:related}) and conclude (Sect.~\ref{sec:conclude}).

\section{Background: Neural Clause Guidance in Vampire}
\label{sec:background}

In this work, we conduct new experiments with a neural clause selection guidance system for the ATP \vampire{} \cite{DBLP:conf/cav/BartekBCHHHKRRRSSV25} introduced last year \cite{DBLP:conf/cade/Suda25}.
The system uses a single clause selection queue ordered by scores assigned to clauses by a neural network trained on past successful proof attempts.
The network consists of several connected computational blocks:
\begin{itemize}
\item
	a graph neural network (GNN) \cite{DBLP:conf/iclr/KipfW17,DBLP:conf/nips/HamiltonYL17} that processes the input problem's
	clause normal form (CNF) in several message-passing rounds and produces vectorial representations (called \emph{embeddings})
	of the input clauses and of the problem's signature symbols,
	

\item
	a recursive neural network (RvNN) \cite{DBLP:conf/ki/KuchlerG96} unfolding along the clause derivation history;
	i.e., using the GNN-produced input clause embeddings as a base case, computing an embedding of
	each derived clause by combining embeddings of its parents with information about the derivation rule used,
	
	
\item
	a second RvNN running along \vampire{}'s (perfectly-shared) term structure, using the GNN embeddings of the
	signature symbols as a base case and computing an embedding of each
	compound term by combining embeddings of constituent sub-terms with that of the leading symbol, and

	
\item
 	a fully-connected NN with one hidden layer that processes the concatenation of the two embedding kinds
	(clause derivation history embedding and the clause term structure embedding) together with several
	simple clause features (age, weight, number of literals, etc.) to a single clause score.\footnote{The contribution of these simple features
	is not big but still measurable (cf. Fig.~4 of \cite{DBLP:conf/cade/Suda25}).}

\end{itemize}
While the GNN runs only once, after the input problem preprocessing yields the corresponding CNF,
the computation of the two RvNNs is interleaved with the saturation process and new clauses are evaluated as they are getting derived.

We start the training process with a randomly initialized network and repeat the following
sequence of steps until the performance stops improving:
\begin{itemize}
\item
	Evaluate (plain or neurally-guided) \vampire{} on the training problems and collect
	\emph{traces} from the successfully solved ones. A trace records the clause traffic
	in the prover, so that for any moment in which a clause selection decision was made,
	we can reconstruct the candidate clauses there (i.e., the content of the passive set)
	and which of these candidates later ended up in the found proof.

\item
	Use the collected traces to compute a loss function expressing
	``in each clause selection step, the score of proof clauses should be better than that of the remaining ones''
	and update the network parameters using several gradient descent steps.

\end{itemize}
Since the randomly initialized network at the beginning is not very good at guidance, in the first loop iteration
we instead collect traces using the standard (built-in) clause selection heuristic. We also remark that both
the prover evaluation and network updates are computationally relatively expensive operations,
but can be efficiently parallelized.

We refer the reader to our previous paper for additional details \cite{DBLP:conf/cade/Suda25}.\footnote{
Several small differences to the exact setup used in \cite{DBLP:conf/cade/Suda25} are detailed in Appendix~\ref{sec:differences}.
}

\section{Four Datasets} 
\label{sec:datasets}

We use the following four problems sets (benchmarks), later referred to as \emph{datasets}, for our experiments. Note that all of them come in the TPTP format and are thus directly accessible to \vampire{}.

\paragraph{TPTP:} Starting from the CNF, FOF, and TF0 format problems of the TPTP library version 9.1.0 \cite{Sut17},
we excluded non-theorems and problems containing arithmetic. This left us with \num{15500} problems in this dataset. 
Our previous work \cite{DBLP:conf/cade/Suda25} uses TPTP v9.0.0 and does not exclude the satisfiable problems.
This makes the ``percentage library solved'' values incompatible,
but the ``percentage improvement'' values remain comparable (modulo the library version).

\paragraph{Mizar40:} This refers to \num{57880} ``bushy'' problems \cite{DBLP:journals/jar/KaliszykU15a} from the TPTP export of the Mizar Mathematical Library \cite{DBLP:journals/jfrea/GrabowskiKN10}
using the MPTP infrastructure \cite{DBLP:journals/jar/Urban06}.
Here ``bushy'' \cite{Urban_MPTPChallenge} means that each conjecture is only accompanied by those previously occurring lemmas and theorems that are needed (according to the original human-supplied proof in the library)
to prove the conjecture.\footnote{In contrast, ``chainy'' (hammer) \cite{Urban_MPTPChallenge} versions of problems contain all the previously occurring theory development and typically are much harder to solve automatically.}
\mizar{} has been used by numerous ML-for-ATP evaluations before \cite{DBLP:conf/itp/JakubuvU19,DBLP:conf/frocos/Suda21,DBLP:conf/lpar/ChvalovskyKPU23}.

\paragraph{Isabelle:} Refers to \num{276363} Isabelle Sledgehammer \cite{DBLP:conf/cade/BohmeN10} problems extracted by Goertzel et al.~\cite{DBLP:conf/itp/GoertzelJKOPU22}
using the Isabelle tool Mirabelle. Out of the two encodings provided, we picked the many-sorted first-order (TFF) encoding (the content of folder \texttt{jd\_tff2} in their repository).


\paragraph{CoqHammer:} An analogous tool to Sledgehammer was developed by Czajka and Kaliszyk \cite{DBLP:journals/jar/CzajkaK18} for Coq/Rocq \cite{DBLP:series/txtcs/BertotC04}.
Using the same set of packages as Blaauwbroek et al.~\cite{DBLP:conf/icml/BlaauwbroekORMP24}, we obtained \num{82366} TPTP problems in the FOL fragment.\footnote{
\url{https://github.com/quickbeam123/calibrating-deepire-suppementary-materials} contains links to the datasets
and other information relevant for reproducibility.}



\section{Experimental Setup}
\label{sec:expSetup}

Unless stated otherwise, the experiments reported below evaluate \vampire{} using a limit of \num{32} billion CPU instructions per problem,\footnote{
Instruction limiting leads to more robust results than time limiting in situations (like ours) where many processes compete for the main memory (cf.~Appendix A of \cite{EasyChair:7719}).}
which roughly amounts to \SI{12}{\second}\footnote{Low time limits correspond to the hammer settings in which we are interested here.} time on our servers.
These are equipped with AMD EPYC 7513 (128 cores with \SI{2.6}{\giga\hertz}) and \SI{500}{\giga\byte} RAM
and run Ubuntu 22 (kernel 5.15.0-161-generic).

We always evaluate \vampire{} using 120 of the 128 available cores. It is worth pointing out that each successful run must be followed up by
a trace collecting rerun, which uses an increased time and instruction limit, because trace collecting is more expensive than pure proof search.
When training models, we enable input and internal shuffling \cite{DBLP:conf/cade/Suda22} to solve more problems and diversify the traces,
so we need to remember and reuse the same random seed for the trace collecting run. However, final model evaluation is done without shuffling.

Training is done in parallel using 64 cores only, so that the computation graphs comfortably fit into the available RAM.
The parallel training setup is inherited from our previous work \cite{DBLP:conf/frocos/Suda21} and involves non-sequential updates
of the trained model which works as a form of regularization (cf. Sect.~3.4 there).

We alternate prover evaluation (and trace collection) with model training for a total of 25 iterations.
On \tptp{}, a typical iteration takes around \SI{90}{\minute} and thus the whole model tuning experiment
takes up approximately \num{1.5} days there. Each trained model is $\sim$\SI{1}{\mega\byte} in file size, 
which corresponds to approximately 0.25 million neural parameters (weights).

\section{Per-Dataset Neural Boosting and Transfer Learning}
\label{sec:whereDefaultModelGetsBorn}

As a first experiment, we take our four datasets and train a neural clause selection guidance model on each.
We always use \vampire{}'s default strategy\footnote{This is one fixed strategy with a good performance,
which \vampire{} uses when nothing else is specified; i.e., when invoked simply via \texttt{./vampire <problem>}.} as the basis, to provide traces for the first training iteration
and to fix the prover configuration throughout the whole training and evaluation process.

We trained the four respective guiding models on 1) the full \tptp{}, 2) 15 thousand randomly selected problems from
\mizar{} and \isabelle{}, leaving 5 thousand different problems aside for testing, and 3)
30 thousand randomly selected problems from \coqhammer{}.\footnote{
We used a train set twice as large here, compared to \mizar{} and \isabelle{},
because the default strategy, from which we learn, solves so few problems on \coqhammer{}.
Doubling the train set size also meant correspondingly larger evaluation times.} For the last dataset,
we followed Blaauwbroek et al. \cite{DBLP:conf/icml/BlaauwbroekORMP24} and picked
5 thousand test problems from their list of separate test Coq packages (distinct theory developments), so there cannot be any
interleaving of related theorems and lemmas between the train and test problems.\footnote{This implements ``no learning from the future'', i.e.,
the lemmas and symbols introduced in the test packages are not seen during the training.}

We did not leave any \tptp{} problems for testing, not to reduce its relatively small size further.
Note, however, that a test set improvement of \SI{20}{\percent} was already established in our previous work \cite{DBLP:conf/cade/Suda25}.
Moreover, it can be argued that it is only thanks to generalization (which a train/test split normally tries to establish)
that a neural model can keep improving and solving new (training) problems in each iteration.
Viewed this way, a separate train/test split is not strictly necessary.

%
%
%

\subsection{Boostability}

\begin{table}[t]
    \centering
    \caption{Boosting the performance 
    by neural clause selection guidance on our four datasets.
      The used format is $(\mathit{default} \rightarrow \mathit{boosted})/\mathit{total}$ and refers to the number of problems solved
      and the size of the corresponding set (train or test), respectively. 
      }
    \label{tab:boostPerDataset}
    \setlength{\tabcolsep}{3pt}
    \begin{tabular}{l|rr|rr}
dataset & \multicolumn{2}{c|}{train} & \multicolumn{2}{c}{test} \\
\hline
\tptp{} & $(\num{8511} \rightarrow \num{10864})/\num{15500}$ & (+\SI{27.6}{\percent}) & \multicolumn{2}{c}{---}  \\
\mizar{} & $(\num{4733} \rightarrow \num{10098})/\num{15000}$ & (+\SI{113.4}{\percent}) & $(\num{1543} \rightarrow \num{3101})/\num{5000}$ & (+\SI{101.0}{\percent})  \\
\isabelle{} & $(\num{6435} \rightarrow \num{8788})/\num{15000}$ & (+\SI{36.6}{\percent})&  $(\num{2178} \rightarrow \num{2715})/\num{5000}$ & (+\SI{24.7}{\percent}) \\
\coqhammer{} & $(\num{2813} \rightarrow \num{9799})/\num{30000}$& (+\SI{248.3}{\percent})& $(\num{399} \rightarrow \num{772})/\num{5000}$  & (+\SI{93.5}{\percent})  \\
    \end{tabular}
\end{table}

In Table~\ref{tab:boostPerDataset}, we can see how training a guiding model can improve \vampire{}'s performance on
the individual datasets. Notice that the hardest \coqhammer{} dataset benefits the most (boost +\SI{248.3}{\percent}),
although not quite as much of its train boost (compared to \mizar{} and \isabelle{}) persists to the unseen test set (+\SI{93.5}{\percent}).
The reason could be the stricter train/test-split hygiene used with \coqhammer{}.

While the test performance is always smaller than the train performance, even the test performance boost is
substantial across all datasets, with interesting differences between the datasets.
Notably, the \isabelle{} dataset appears to be the hardest to record an improvement on.

\begin{figure}
    \centering
    \includegraphics[scale=0.7]{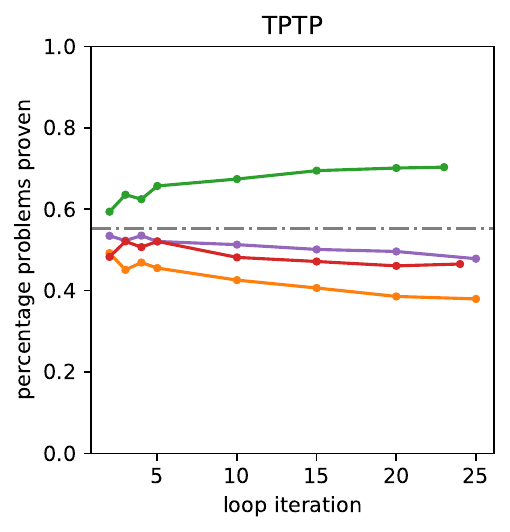}
    \includegraphics[scale=0.7]{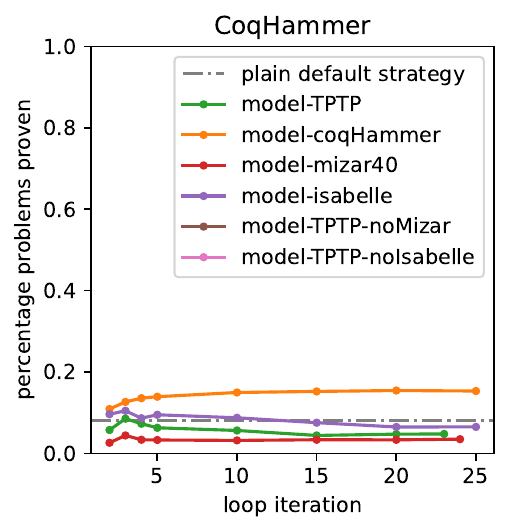}
    \includegraphics[scale=0.7]{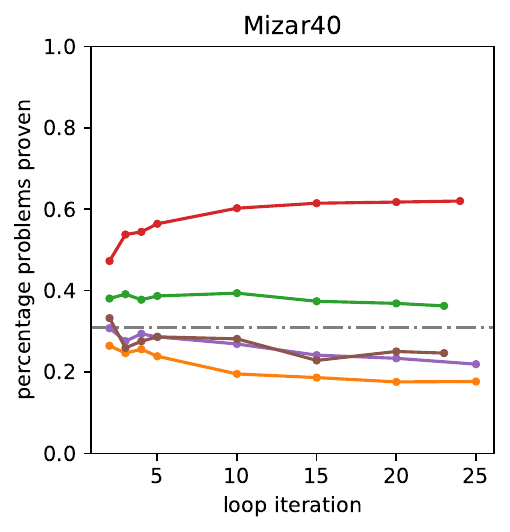}
    \includegraphics[scale=0.7]{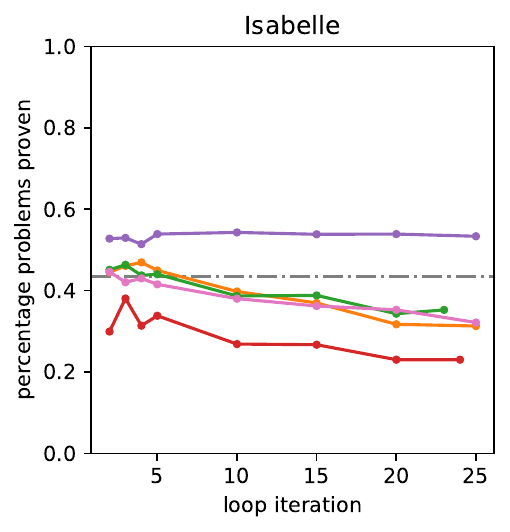}
    \caption{
    Performance evaluation---across our datasets---of  strategies boosted by models trained on the individual datasets.
    Except for \tptp{}, which does not have a test split, the evaluation is on test problems.
    The $x$-axis follows the evolution of the models during training.
    The legend at the upper right corner (CoqHammer) is meant to be shared.}
  \label{fig:nontransfer}
\end{figure}

\subsection{Knowledge Transfer}

While Table~\ref{tab:boostPerDataset} only reveals the performance of the best model (picked according to the training set),
Fig.~\ref{fig:nontransfer} offers a more refined perspective. First, instead of raw problem counts, it uses percentages.
Moreover, the $x$-dimension in each of the four sub-figures tracks the evolution of each model through its training process.
Namely, models at iterations 2--5,10,15,20, and the iteration with the best training performance (always >20) are highlighted.
Most importantly, however, each of the four sub-figures also tracks the models trained on the other datasets.

It can be seen from Fig.~\ref{fig:nontransfer} that as a model specializes to its own training dataset and improves on it,
it also gets worse on any of the other datasets to the point of consistently scoring worse than the default (unguided) strategy there.
In more detail, there seems to be a single exception, a model trained on \tptp{} improves over the default strategy (always denoted by the gray dashed line) on the \mizar{} problems (Fig.~\ref{fig:nontransfer}, bottom left).
However, the TPTP library itself contains \num{1840} problems exported from Mizar by the MPTP tool. When these are excluded and a separate model 
is trained on the reduced set \dataset{TPTP-noMizar}, its performance also soon drops below that of the default strategy.
(Similarly, the \isabelle{} subfigure shows the performance development of a model trained on \dataset{TPTP-noIsabelle},
which is \tptp{} without \num{1276} problems mentioning ``Sledgehammer'' in their source field.)

This is a clear negative answer to the question about knowledge transfer. We were a bit surprised 
by the clarity of the trend, especially with \tptp{}, which is known to contain problems 
from many diverse sources (besides Mizar and Isabelle) collected over several decades.
The observation, however, also offers a hint as to what our guiding networks 
could be capitalizing on to achieve better performance. It is likely more of ``make sure problem
encoding specifics do not adversely affect proof search'' and less of learning how a superposition-based saturation
prover should be controlled in general.

For instance, the \isabelle{} and \mizar{} TPTP encodings we use are quite different (type \emph{guards} in Mizar vs type \emph{tags} in Isabelle),
resulting in likely very different initial clausal (CNF) graphs processed by the GNN to obtain the initial embeddings of symbols and clauses. One plausible explanation of the lack of transfer is thus the major (even if perhaps a bit superficial from the ATP point of view) difference in the graphs that the GNN is trained and evaluated on.
We started looking at individual problems, how different models score their clauses, and how these scores impacts proof search.\footnote{
See Appendix~\ref{sec:shadow} for such an analysis on several problems.}
This painstaking analysis has so far not yielded a single dominant explanation. 
We leave a more systematic investigation for future work.

The classical machine learning remedy to this ``eager adaptation'' should be to reduce the model's capacity,
but while a smaller network would very likely become harder to train (and thus to overtrain), it is not guaranteed that it would start picking up ``deeper patterns''
(that go beyond exploiting the used encodings).
The more modern, ``deep learning''-era, advice is to simply get more data from all possible sources of interest. In this light, \tptp{} with its several decades of history 
and more than 15 thousand first-order problems appears still too small.

\subsection{Overfitting To Training Data}

In a last glimpse at Fig.~\ref{fig:nontransfer}, we can observe that not all four datasets agree on how long it is beneficial
to train a model for the benefit of its test performance.

Setting aside the TPTP subfigure, where we are actually observing the train performance,
we see that on \mizar{}, the test performance kept improving through the whole 25 iterations.
On the other hand, the test performance on \isabelle{} is fairly constant already since iteration 5 and
actually maxes out at iteration 10. (With \coqhammer{}, we see a similar plateau maxing at iteration 20).
While the first behavior indicates good generalization, the latter points in the direction of overfitting.
In the case of \isabelle{} we could try mitigating this by training on more than just
15 thousand out of the almost quarter of a million available problems and paying for it with a higher computational cost,
which we intend to do as part of future work.

\subsection{Training on All Four Datasets at Once}

By putting all the four training datasets on one large pile (\num{15500} problems from \tptp{}, \num{15000} from \mizar{}, \num{15000} from \isabelle{}, and \num{30000} from \coqhammer{})
and waiting for 17 days for the 25 iterations of the training process to finish, 
we were able to obtain a model that is almost as good as 
the ones dedicated to the individual datasets we talked about until now.
In particular, the model can reach (when comparing test performance of the best iteration on each side) 
\SI{100.0}{\percent} performance of the dedicated \tptp{} model on \tptp{},
\SI{99.4}{\percent} performance of the dedicated \mizar{} model on \mizar{},
\SI{101.4}{\percent} performance of the dedicated \isabelle{} model on \isabelle{},
and \SI{95.3}{\percent} performance of the dedicated \coqhammer{} model on \coqhammer{}.

This result suggests that 1) the capacity of the model is not (at the scales discussed in this paper)
the main limiting factor of its capabilities and 2) the GNN at the bottom of the model's architecture
is likely very good at recognising various input sub-classes (such as the distinct encodings used by our four datasets,
but also---at a lower granularity---within each dataset) so that the subsequent scoring
of the clauses may become sub-class dependent.

\section{Strategy Boosting}
\label{sec:strategies}

\begin{figure}[t]
    \centering
    \begin{minipage}{0.97\linewidth}
\begin{lstlisting}[style=mystyle]
def hillClimbingGreed(all_problems, seed_strategy = default):
  cur_probs = all_problems
  cur_best = seed_strategy
  evaluate(seed_strategy,all_problems)
  for i in range(5):
    cur_best = best strategy on cur_probs evaluated so far
    cur_best_score = evaluate(cur_best,cur_probs)
    for j in range(3):
      for opt, values in options_to_vary:
        for val in values:
          score = evaluate(cur_best[opt:=val],cur_probs)
          if score > cur_best_score:
            cur_best_score = score
            cur_best = cur_best[opt:=val]
    print("Champion of iteration",i,"is",cur_best)
    cur_probs -= problems solved by cur_best
\end{lstlisting}
    \end{minipage}
    \caption{Our algorithm for constructing small sets of complementary strategies. Given a problem set \texttt{all\_problems} 
    to optimize for and a strategy \texttt{seed\_strategy} to start with (naturally initialized as the prover's \texttt{default} strategy),
     the algorithm calls the function \texttt{evaluate} to compute the number of problems
    solved by a given strategy on a given set of problems (under an implicit fixed time limit). The results of calls to \texttt{evaluate}
    are cached, 
    which makes picking the current best strategy on line 6 computationally cheap.
    The outer loop starting at line 5 governs the greedy cover and here hardcodes 5 as the number of complementary strategies to produce.
    The inner loops attempt to improve the current best strategy by always picking an option to vary, trying out all available values for it 
    and committing to the one with the best score. Each option is tried up to 3 times. In an actual implementation, the loop of line 8
    can be broken out of early if \texttt{cur\_best} does not change for a whole iteration.
} 
    \label{fig:hillClimbingGreed}
\end{figure}

An ATP strategy is a fixed configuration of the prover, defined by
choices of preprocessing steps, enabled inference and simplification
rules, term ordering, literal selection, and other heuristic parameters.
It is well known that combining multiple strategies into sequences (called \emph{schedules} or \emph{portfolios}),
executed either sequentially or in parallel, can greatly improve the prover performance in practice.
Informally, the desired property of strategies is their \emph{complementarity},
which means that each strategy is good at solving different kinds of problems.

In this section, we investigate how neural guidance interacts with proving strategies and strategy schedules.
We focus on the \tptp{} dataset and unless specified otherwise we keep the established instruction limit of \SI{32000}{Mi} for our evaluations.

\subsection{Portfolio Construction via Greedy Cover and Local Search}
\label{subsect:hillClimbGreed}

We start by constructing a small baseline set of complementary strategies. To this end, we implement
an algorithm combining a standard greedy cover approach---repeatedly picking the strategy that solves
the largest number of currently unsolved problems---with hill climbing
to locally improve the current best candidate (see  Fig.~\ref{fig:hillClimbingGreed}).
A notable aspect of this algorithm is that, with a sufficiently reduced time limit and enough parallelization,
it is feasible to run the algorithm to completion over \vampire{}'s more than 80 main configuration options---many of which are multi-valued rather
than binary\footnote{For options with a potentially unbounded set of values, such as the age-weight ratio,
  we hand-picked small representative sets of values.}---within
approximately one day. To obtain the strategies mentioned below
we used an instruction limit of \SI{8000}{Gi} and a random subset of TPTP of size 7680.

In contrast to several previous approaches to strategy invention, such as BliStr \cite{DBLP:conf/gcai/Urban15} or HOS-ML \cite{DBLP:conf/mkm/HoldenK21},
which cluster problems based on similar performance and save computation by optimizing strategies only on the clusters,
our less sophisticated algorithm delivers strategies optimized globally, whose mutual
relationship is thus arguably easier to interpret.

\paragraph{Five Baseline Strategies to Boost}
The five baseline strategies our 
algorithm selected\footnote{See Appendix~\ref{sec:strategiesTPTP} for their listing.} differ substantially from \vampire{}'s default
strategy. In particular, none of them uses a complete literal selection function \cite{DBLP:conf/cade/HoderR0V16}. 
The first strategy uses Twee's \emph{goal transformation} \cite{DBLP:conf/cade/Smallbone21},
which (together with several other non-default option settings) contributes to its markedly stronger performance compared to  \vampire{}'s default strategy (9265 vs 8574 problems solved).\footnote{
The reported 8574 problems were solved by the default strategy
without the input and internal shuffling,
which is, as mentioned, turned on during training and lead to the slightly lower value of 8511
reported in Table~\ref{tab:boostPerDataset}.}
As for complementarity, we were surprised to never see the Discount 
loop being used nor the AVATAR \cite{DBLP:conf/cav/Voronkov14} architecture getting turned off,
although both settings are known to lead to many unique solutions when varied in isolation.

\subsection{Boosting the Baseline Strategies Individually}
\label{subsec:fiveStratBoost}

We trained a separate guiding model for each of the five strategies. This can be viewed as aiming to obtain expert models that specialize (i) primarily to the specific variant of the superposition calculus employed,\footnote{Recall that the network can
discriminate individual inference rules and adapt to how useful they are for producing a proof clause in specific contexts. 
This also means that a model trained only on runs never using a particular inference rule $R$ will likely behave poorly when
employed to guide a strategy using $R$ a lot.}
and (ii) secondarily to the types of benchmark problems that the strategy tends to solve.
We then evaluated each boosted strategy
in isolation and also their (incremental) union when following the original greedy sequence.

The obtained results are shown in Table~\ref{tab:greedyBoost}.
%
\begin{table}[t]
    \centering
    \caption{Performance of five complementary strategies 1--5 (left) and their neurally-boosted counterparts $1'\text{--}5'$ (right).
    We show how each subsequent strategy contributes to the greedy cover (contrib.) and also its plain
    performance (\#solved). }
    \label{tab:greedyBoost}
    \begin{tabular}{rr@{\hspace{10pt}}|@{\hspace{10pt}}rcr@{\hspace{10pt}}|@{\hspace{10pt}}rr}
\multicolumn{3}{r}{plain proving strategies} &|& \multicolumn{3}{l}{neurally-boosted counterparts} \\
id & contrib. & \#solved & (boost by) & \#solved & contrib. & id \\
    \hline
1 & \num{9265} & \num{9265} & +\SI{21.6}{\percent} & \num{11264} & \num{11264} & ${1}'$ \\
2 & \num{826} & \num{8395} & +\SI{29.1}{\percent} & \num{10839} & \num{434} & ${2}'$ \\
3 & \num{298} & \num{8999} & +\SI{18.6}{\percent} & \num{10675} & \num{98} & ${3}'$ \\
4 & \num{243} & \num{5936} & +\SI{19.9}{\percent} & \num{7118} & \num{338} & ${4}'$ \\
5 & \num{87} & \num{8078} & +\SI{30.4}{\percent} & \num{10533} & \num{73} & ${5}'$ \\
\hline
union:        & \num{10719} & $\longrightarrow$ & +\SI{13.9}{\percent}  & $\longrightarrow$ & \num{12207} \\
  & (\SI{15.7}{\percent} of 1) &&&& (\SI{8.4}{\percent} of $1'$)
    \end{tabular}
\end{table}
%
%
Each of the 5 strategies gets on average boosted by \SI{24.0}{\percent} by its trained neural guidance model.
However, the strategies (unsurprisingly) do not grow just in complementary directions:
the performance of the champion strategy 1 is getting improved by \SI{15.7}{\percent} when complemented by 2--5,\footnote{Such a portfolio would require 5 times more time to run to completion.
Although, typically, most of the combined performance of several strategies can be reaped much earlier by a more refined strategy schedule \cite{DBLP:conf/ijcar/BartekCS24}.}
while in the neurally-boosted version 1' gets ``only'' improved by \SI{8.4}{\percent} by also considering $2'\text{--}5'$.

The table uses the same order both on the left and on the right to show the cumulative contribution of each strategy. This leads to a non-monotonic progression
for the neurally-boosted strategies, where, e.g., strategy 3' only contributes 98 problems while 4' brings 338 additional ones. Strategy 4/4'
is the only one in the sequence enabling the SInE premise selection heuristic \cite{DBLP:conf/cade/HoderV11}, which is likely the main reason
why it brings many new problems even after boosting. At the same time, complementarities of the strategies
in how they configure the standard clause selection heuristic---for which \vampire{} has several options---are most likely vanishing
with the boosting, because neurally guided \vampire{} ignores these options (as it delegates clause selection fully to the model).
%
In total, we observed a \SI{13.9}{\percent} improvement of the 5-strategy portfolio's performance through individual strategy boosting.

\subsection{Boosting All Baseline Strategies Jointly}
\label{subsect:singleGuidingModel}

It is not necessary to train 5 individual models to boost 5 strategies. By slightly modifying the training scripts, we were able to train a single model
(of the same capacity) on traces generated by all five strategies at once. The training process slows down correspondingly, but the final model is almost
as good as the 5 individual ones.
The average strategy boost is by \SI{23.8}{\percent} 
and the combined coverage of the five single-model-boosted strategies 1--5 is \num{12195} problems.

\subsection{Making Vampire Smart Again}
\label{subsect:smartAgain}

A specific strategy we were especially keen to test boosting is one which replaces the standard ordered resolution and superposition calculus
(governed by a heuristically selected simplification ordering)
with general (unordered) resolution and plain paramodulation (no ordering constraints). We were inspired for this by the work of Goertzel \cite{DBLP:conf/cade/Goertzel20}
who tried the analogous experiment in the context of E prover \cite{DBLP:conf/cade/0001CV19} and ENIGMA clause-selection guidance \cite{DBLP:conf/cade/ChvalovskyJ0U19}.

The idea is that an unconstrained calculus is much more prolific and often causes the prover to choke before a proof can be found.
The research question is then whether a well-trained clause-selection guidance can compensate for such inefficiency.
In our case, the unconstrained strategy $S_0$\footnote{Can be invoked via \texttt{./vampire -s 0 -to incomp <problem>}.}
solves \SI{40.4}{\percent} of TPTP
(compared to the default's \SI{55.3}{\percent}) 
and reaches \SI{61.3}{\percent} after boosting. This is less than the boosted default's \SI{70.3}{\percent}, but quite a bit more than the performance of the plain default.
This answers the question affirmatively.

%

Strategy $S_0$ is noteworthy also from the perspective of proof
complexity. It is known that certain formulas admit short general
resolution proofs but only long ordered-resolution proofs; analogous
phenomena arise when comparing plain paramodulation with superposition
under ordering restrictions. We could, however, not confirm that this is of practical concern
on TPTP. Although the boosted version of $S_0$ solved several hard problems\footnote{Namely the set theory problems \probtptp{SET276-6},
\probtptp{SET277-6}, \probtptp{SET289-6}, and \probtptp{NUM247-1} (the last on ordinal numbers, but still essentially set-theorethical), all of TPTP rating 1.0 and TPTP status \texttt{Unknown}.}
and was the only strategy claiming their solution for a long time during our 20 day Spider-style strategy search experiment (see Sect.~\ref{subsect:spiderStyleStrategySearch} below),
these problems were later solved also by a different strategy employing the standard superposition calculus (constrained by a simplification ordering).

\subsection{Spider-style Strategy Search}
\label{subsect:spiderStyleStrategySearch}


More than five complementary strategies are needed to construct a strong proving portfolio such as the one \vampire{} employs during CASC \cite{Sut16}.
In spider-style strategy search \cite{DBLP:conf/ijcar/BartekCS24}, strategies are discovered by iteratively attempting to solve 
a single as-of-yet unsolved problem using a randomly sampled strategy, and, on success, by subsequently optimizing
the random strategy on that problem by local search through the option space. An optimized strategy is then evaluated on the whole
problem set, to establish which problems get newly covered.

\begin{figure}[t]
    \centering
    \includegraphics[scale=0.7]{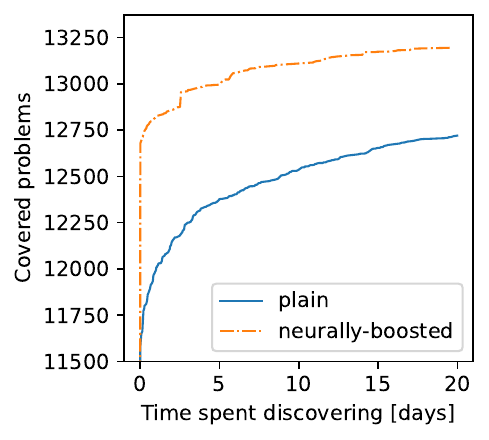}
    \includegraphics[scale=0.7]{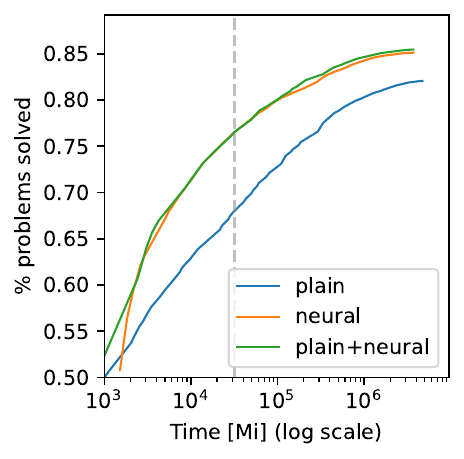}
    \caption{Left: Cumulative problem coverage in time, growing as spider-style strategy search keeps discovering new strategies. Right: Cumulative performance of a plain, a purely neural, and a combined greedy strategy schedules (mark at \SI{32000}{Mi}). A schedule could employ any strategy from its respective pool (plain, neural, or both) as filled during the preceding 20 days.}
    \label{fig:strategiesAndSchedules}
\end{figure}

On the TPTP library, this process may continue covering new problems even after 20 days of search, albeit with diminishing returns \cite{DBLP:conf/ijcar/BartekCS24}.
In Fig.~\ref{fig:strategiesAndSchedules} (left), we can see a visualization of a problem covering process that we ran to obtain strategies---both plain and neurally-boosted---for the purpose
of establishing the value of neural guidance for the power of the prover ``in the limit''. We seeded the process (at time 0) the following way:
\begin{description}
\item[plain:] Using the \texttt{hillClimbingGreed} algorithm (Sect.~\ref{subsect:hillClimbGreed}), we generated 10 (rather than 5) complementary strategies
 and evaluated them under an instruction limit of \SI{64000}{Mi}.
\item[neurally-boosted:] Analogously, to obtain strong neurally-boosted strategies, we evaluated strategies $1'\text{--}5'$ of Table~\ref{tab:greedyBoost},
i.e.~the first 5 strategies from \texttt{hillClimbingGreed} each with its own guiding model, plus the default strategy with its guiding model (introduced in Sect.~\ref{sec:whereDefaultModelGetsBorn}),
plus the model trained jointly on all the first 5 strategies (as explained in Sect.~\ref{subsect:singleGuidingModel}) evaluated again on all 5. Also these 11 mentioned strategies were run for \SI{64000}{Mi} each.
\end{description}
After this initialization phase, the timer was started and two servers independently performed spider-style strategy search for 20 days---one in the plain setting and one in the neurally boosted setting. By the end of this period, the plain search had produced \num{569} strategies covering \num{12720} problems, while the neurally boosted search had produced \num{244} strategies covering \num{13195} problems. Taken together, the two searches covered \num{13247} distinct problems.




\subsection{Greedy Portfolio Construction}
\label{subsect:greedyPortfolio}

Our previous work \cite{DBLP:conf/ijcar/BartekCS24} presents a simple greedy algorithm for constructing a strategy schedule
using the information about which problems are getting solved by which strategy at what moment in time. By a schedule
we mean an assignment of time budgets to every given strategy and specifying their execution order.\footnote{
So the algorithm must do more than just compute the greedy cover sequence of the set of solved problems, such as the one presented in Table~\ref{tab:greedyBoost}.}
If we ask what the performance of such a schedule would be when constructed from
1) just our plain strategies, 2) just the neurally-boosted ones, and 3) both kinds of strategies combined, we get the cactus plot in Fig.~\ref{fig:strategiesAndSchedules} (right).\footnote{
Technically, the plot actually comes from the budget-less version of our algorithm, recording performances of gradually (greedily) extended schedules
as they are getting produced (see Sect.~5.1 of \cite{DBLP:conf/ijcar/BartekCS24} for details).}

%
%

The combined schedule covers \SI{4.1}{\percent} more problems than the plain one. 
We see that when we shift our focus towards the limit of what is provable by \vampire{},
neural boosting does not add quite as much as in the single-strategy setting.
However, the ultimate coverage is not necessarily the most relevant factor for the users.
The combined schedule only needs \SI{3.7}{Ti} (roughly 0.5 hours single-core execution), compared to \SI{4.7}{Ti} for the plain schedule.
Also, if we focus on the schedules' performance at our standard \SI{32000}{Mi} mark, we
get \SI{76.1}{\percent} versus \SI{67.8}{\percent} problems covered, which is a much more substantial \SI{12.2}{\percent} improvement.

\section{Strategies versus Neural Boosting}
\label{sec:stratsAndBoosting}


Both neural clause selection guidance and the hill climbing strategy optimization
seek to improve the prover performance in their own way. Thinking back
of the experiment in Sect.~\ref{sec:whereDefaultModelGetsBorn}, we may now ask
questions such as:
What is the ultimate performance achievable by a single strategy that is both locally optimized and neurally boosted?
To what degree did neural boosting only compensate for the inefficiencies of the default
strategy on each individual dataset? And, finally,
how much do strategies tuned for a particular dataset perform on a different one?

\begin{figure}
    \centering
    \includegraphics[scale=0.7]{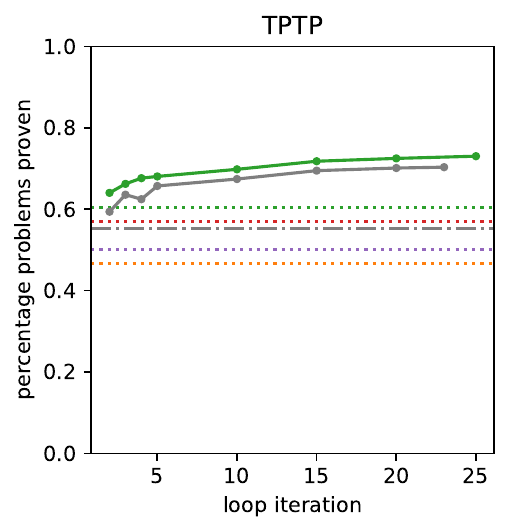}
    \includegraphics[scale=0.7]{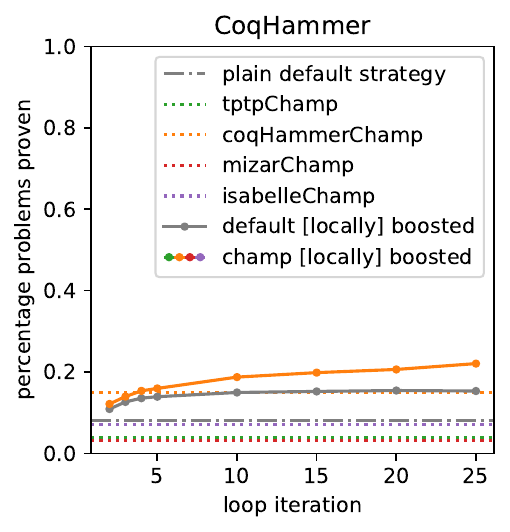}
    \includegraphics[scale=0.7]{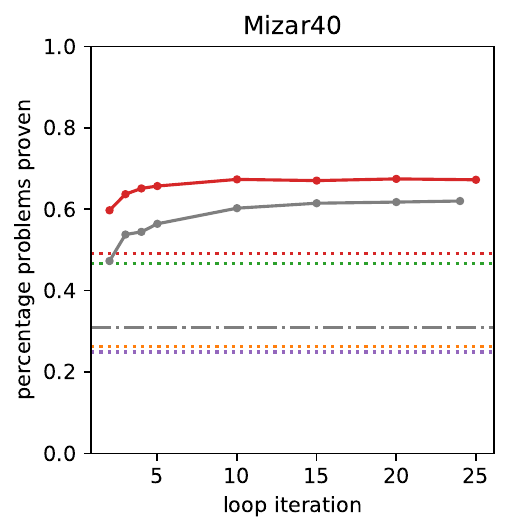}
    \includegraphics[scale=0.7]{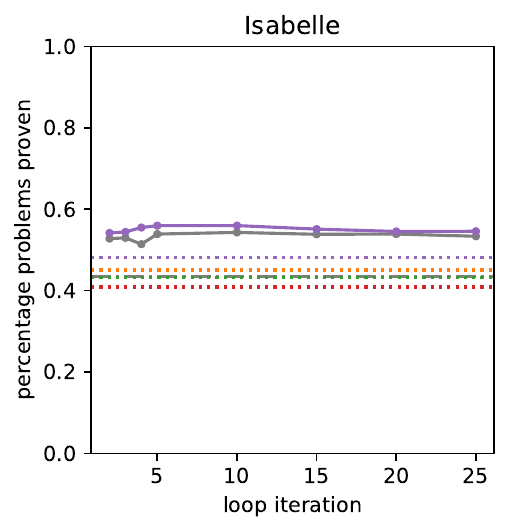}
    \caption{Per-dataset performance comparison of champion strategies (dotted, in color), which are the \texttt{hillClimbingGreed} winners on the individual datasets,
    and their neurally-boosted analogues (solid, in color). The axis semantics are the same as in Fig.~\ref{fig:nontransfer}.
    Also the default strategy's performance (dashed, gray) and its local boosting (solid, gray) are taken from Fig.~\ref{fig:nontransfer}.}
    \label{fig:champs}
\end{figure}

Some answers to these questions can be found in Fig.~\ref{fig:champs},
which mirrors Fig.~\ref{fig:nontransfer} of Sect.~\ref{sec:whereDefaultModelGetsBorn} in its layout.
For reference Fig.~\ref{fig:champs} again shows the performance of the default strategy (dashed, gray horizontal lines) on each dataset
and of its neurally boosted counterpart trained on that dataset (gray curves).
In addition, it shows the performance of four locally optimized champion strategies (dotted),
each being a ``Champion of iteration 1'' of the \texttt{hillClimbingGreed} run on one of the four datasets.
Finally, the performance progression rendered in color is, in each subfigure, the local champion strategy as it gets neurally boosted.

Perhaps unsurprisingly, it is true across all four datasets that boosting a local champion gives us the ultimate performance,
although the relative benefit differs per dataset. Regarding the champion strategies,\footnote{Their list can be found in Appendix~\ref{sec:champStrats}.}
we can say that the \tptp{} and \mizar{} champions and the \coqhammer{} and \isabelle{} ones behave similarly:
On \tptp{} and \mizar{} both the \tptp{} and \mizar{} champions improve over the default strategy,
while the \coqhammer{} and \isabelle{} are worse than the default. Almost a mirror image of this trend can be seen on the \coqhammer{} and \isabelle{} side.
The exception being the \coqhammer{} dataset on which even the \isabelle{} champion fares worse than the default strategy (although only by a little).

It is not immediately clear what the main reasons are behind these similarities and differences. However, 
we noticed that the already mentioned Twee's \emph{goal transformation} \cite{DBLP:conf/cade/Smallbone21}
is shared exactly by the \tptp{} and \mizar{} champions, whereas the \coqhammer{} and \isabelle{} champions share a reservation
towards (and turn off) \vampire{}'s preprocessing step \emph{function definition elimination}, which eliminates definitions by inlining them before saturation starts,
and they also both prefer high-age \emph{age-weight} ratio (4:1 for the \coqhammer{} and 16:1 for the \isabelle{} champion),
i.e., these two champions very much prefer a breadth-first to best-first search approach to clause selection.





Overall, judging by the gap between the weakest and strongest strategies, the \mizar{} dataset exhibits the greatest potential for improvement, both through strategy optimization and neural guidance.
With \tptp{}, roughly half of the library appears relatively easy, while the remaining portion still allows for meaningful performance gains.
In contrast, the remaining two, {ham\-mer-der\-ived} {data\-sets} show a considerably narrower performance spread,
indicating fewer readily solvable instances that could be captured with modest improvements.
This assessment, however, may change in the future as guiding models are trained on larger portions of these comparatively large datasets.

\section{Related Work}
\label{sec:related}


The idea of using successful proofs to improve clause selection through machine learning
methods was discussed in early works by
Schulz~\cite{Schulz1995,DBLP:conf/cade/DenzingerS96,DBLP:conf/ki/Schulz01}. A notable
modern revival was started by the ENIGMA system~\cite{DBLP:conf/mkm/JakubuvU17,DBLP:conf/mkm/JakubuvU18}
building on the E prover \cite{DBLP:conf/cade/0001CV19}. ENIGMA has evolved through multiple iterations that incorporated recursive
NNs~\cite{DBLP:conf/cade/ChvalovskyJ0U19},
GNNs~\cite{DBLP:conf/cade/JakubuvCOP0U20}, and layered architectures~\cite{DBLP:conf/frocos/GoertzelCJOU21}. Similarly, convolutional NNs were explored to guide the
E prover in~\cite{LPAR-21:Deep_Network_Guided_Proof}. These
systems have primarily been trained on Mizar datasets, and later on Isabelle/HOL~\cite{DBLP:conf/itp/GoertzelJKOPU22}.
A large-scale evaluation \cite{DBLP:conf/itp/JakubuvCGKOP00U23} of ENIGMA and related methods reported about 60\% success rate on Mizar40 in the chainy mode and 75\% in the bushy mode in a resource bounded setting comparable to the 2013 Mizar40 results \cite{DBLP:journals/jar/KaliszykU15a}. ENIGMA so far has not used the \emph{dynamic data} collection (introduced in \cite{DBLP:conf/lpar/ChvalovskyKPU23}) and there are further technical differences to the setting used here.

CoqHammer~\cite{DBLP:journals/jar/CzajkaK18} translates Coq theories
to the TPTP format, enabling calls to an external solver. A successful
run of the external solver is used for premise selection to
reconstruct proofs within Coq using the family of \texttt{sauto}
tactics~\cite{czaj20}. Building on the Tactician
platform~\cite{DBLP:conf/mkm/BlaauwbroekUG20},
Graph2Tac~\cite{DBLP:conf/icml/BlaauwbroekORMP24} recommends
appropriate Coq tactics directly. Graph2Tac uses a GNN to build
hierarchical representations for new definitions. Similar to our approach,
the GNN computes embeddings for new symbols. However, it uses
precomputed embeddings for the previously known symbols, in contrast to our method.

NIAGRA~\cite{DBLP:conf/ijcai/FokoueACIKLM023}, a successor of
TRAIL~\cite{DBLP:journals/pami/AbdelazizCMACIK23}, invokes a GNN of the initial CNF to
obtain initial embeddings of symbols, and uses these to seed later clause-local GNN invocations. The authors report
significant success in transfer learning, but only across MPTP,
M2k, and TPTP. However, MPTP and M2k are both Mizar-based datasets. Moreover, their subset of TPTP
is reported to contain fewer than one thousand problems, compared to the more than 15 thousand first-order problems in TPTP version 9.1.0,
and the subset selection criterion is not specified in the work.


\section{Conclusion}
\label{sec:conclude}

We have seen that neural clause selection guidance consistently helps \vampire{}
solve substantially more problems within a given time budget when trained for a particular strategy and on a fixed dataset.
However, this learned advantage disappears when the target dataset is changed,
indicating the network capitalizes mostly on details specific to its training domain rather than
discovering general theorem proving principles or easy-to-transfer math-solving skills.

Also individual proving strategies can be substantially improved by training targeted guidance,
but the added value gets smaller when seen from the perspective of a total coverage of a multi-strategy portfolio.
Still, the observed \SI{12.2}{\percent} improvement on \tptp{} of a portfolio combining both standard and neurally guided strategies
within a practically relevant time budget is more than worth having.

The deep-learning instinct of our era suggest training larger models on larger datasets (we only relied on fractions of our hammer-derived benchmarks here)
and involving the transformer architecture \cite{DBLP:conf/nips/VaswaniSPUJGKP17} with the promise of an emergent intelligent behavior.
A less compute-hungry direction for future research should be a closer look at 
the learned clause/symbol representations and the peculiarities of the used encodings and their comparison across the datasets.





\subsection*{Acknowledgements}
We thank Lasse Blaauwbroek for providing us with the CoqHammer export.
The work was supported by the Czech Science Foundation project 24-12759S
and the ERC grant NextReason (Grant Agreement No.~101200949).



\bibliographystyle{splncs04}
\bibliography{main}

\newpage

\appendix

\section{Technical Differences to Previously Described Setup}
\label{sec:differences}

Compared to our previous work \cite{DBLP:conf/cade/Suda25}, the experiments described in this paper
differ in the following:
\begin{itemize}
\item
  We merged the neural network guidance code with the latest version of \vampire{} (5.0).
  For several important options, \vampire{} 5.0 introduced new default values,
  so that its default strategy should be stronger---at least on \tptp{}---than that of Vampire 4.9 used as the basis previously.

\item
  In previous work, we explained how splitting collected traces into \SI{80}{\percent} training ones
  and \SI{20}{\percent} validation ones, together with the standard early stopping criterion is useful for
  training the model the ``right number of epochs'' in each iteration and thus beneficial for convergence.
  
  Since then we noticed that while this is true for the early iterations, in later ones
  validation loss often was only getting worse, to which our setup responds by terminating after five epochs
  and using a single epoch step to advance the model at least a bit. This results in a lot of wasted training time.
  At the same time, at least the training performance would keep improving anyway.
  
  In the present work, we only use the early stopping criterion for the first two iterations, while for all subsequent ones
  we advance the model by exactly 5 training epochs (using \SI{100}{\percent} of the traces).

\item
  We implemented a new metric called dist\_to\_good, which measures (and averages)
  for every step along the trace, how far on the passive set (normalized between 0.0 and 1.0, using the total size of the passive set)
  would a future proof clause be from getting selected
  if the passive set was sorted by the clause scores obtained by the current model.
  (This metric actually correlates with the loss very well, but has more accessible interpretation.)
  
  We use this metric instead of the differentiable loss as the early stopping criterion in the first two training iterations.

\item
  We use $k=5$ of the GNN message passing rounds instead of $k=8$ in \cite{DBLP:conf/cade/Suda25}.
  This was discovered as sufficient in an ablation experiment \cite{DBLP:journals/corr/abs-2503-07792}. 

\item
  During the training iterations we again exponentially decay the learning rate $\alpha = 0.0002$,
  but with a less aggressive factor of $0.933$ (previously  $0.87055$).
  
\item
  We discovered that our adaptive problem weights (as described in Appendix C of \cite{DBLP:journals/corr/abs-2503-07792}) were buggy,
  in the sense that they actually did not affect the training by modifying the loss. (They were only modifying
  the reported loss in collected logs but did not influence training.) However, a part of this technique (denoted \textit{Boost} in \cite{DBLP:conf/cade/Suda25})
  that works well and that we kept here, is to keep training even from traces discovered longer ago,
  not just the ones from the last evaluation. The magic constant that stayed is 5. After this many iterations a trace becomes obsolete.
  
  In sum, our approach keeps for every problem at most one trace, with a preference for the least recently discovered one (traces are overwritten).
  But will keep learning from a trace of problem $P$ for up to 5 consecutive iterations, even if the problem is not getting solved
  by a newer model anymore.
  
\item 
	We recall that we degrade the limited resource saturation loop into an Otter loop, once a neural guidance kicks in. This was already true before \cite{DBLP:conf/cade/Suda25}. 

\end{itemize}

\section{Five Vampire Strategies Complementary for TPTP}
\label{sec:strategiesTPTP}

{\small
\begin{verbatim}
1:
lrs+1010_1:2_bce=on:bd=preordered:cond=fast:drc=off:fgj=on:lcm=predicate:newcnf=on:
     nm=4:nwc=1.0:sac=on:slsq=on:sp=unary_first:tgt=ground_0

2:
lrs+1011_1:2_drc=off:er=known:fde=unused:fgj=on:nm=16:nwc=2.0:plsq=on:s2a=on:sac=on:
     slsq=on:sp=const_min_0

3:
lrs+1011_1:1_cond=fast:drc=off:er=known:fgj=on:lcm=predicate:nm=4:nwc=1.0:sac=on:
     sfv=off:slsq=on:sp=const_frequency:tgt=ground:to=lpo_0

4:
ott+1011_1:1_bd=preordered:fde=unused:fgj=on:newcnf=on:nm=2:plsq=on:s2a=on:sac=on:
     sp=reverse_frequency:ss=axioms:urr=on_0

5:
lrs-1002_1:10_bce=on:cond=fast:er=known:fde=unused:fgj=on:nwc=2.0:sac=on:slsq=on:
     to=lpo_0
\end{verbatim}
}

\section{Champion Strategies for Our Four Datasets}
\label{sec:champStrats}

{\small
\begin{verbatim}
TPTP:
lrs+1010_1:2_bce=on:bd=preordered:cond=fast:drc=off:fgj=on:lcm=predicate:newcnf=on:
     nm=4:nwc=1.0:sac=on:slsq=on:sp=unary_first:tgt=ground_0

Mizar40:
lrs+1002_3:2_aac=none:add=on:bce=on:cond=fast:eape=off:er=filter:erml=3:fdtod=off:
       ins=6:kws=arity_squared:lma=off:lrd=on:lwlo=on:nm=32:s2agt=16:s2at=1.5:s2pl=no:
     sfv=off:sp=const_max:spb=goal:tgt=ground_0
     
CoqHammer:
dis-1011_4:1_av=off:cond=fast:erd=off:fde=unused:kws=arity_squared:lwlo=on:nm=4:
     sp=arity:updr=off_0
     
Isabelle:
lrs-1002_16:1_aac=none:anc=none:drc=off:fd=preordered:fde=none:kws=arity_squared:
       nm=16:plsq=on:plsql=on:plsqr=1,32:s2pl=no:sac=on:sfv=off:sims=off:urr=on_0
\end{verbatim}
}

\section{Comparing Models Trained on Different Datasets}
\label{sec:shadow}

\begin{figure}
  \centering
  \includegraphics[scale=0.5]{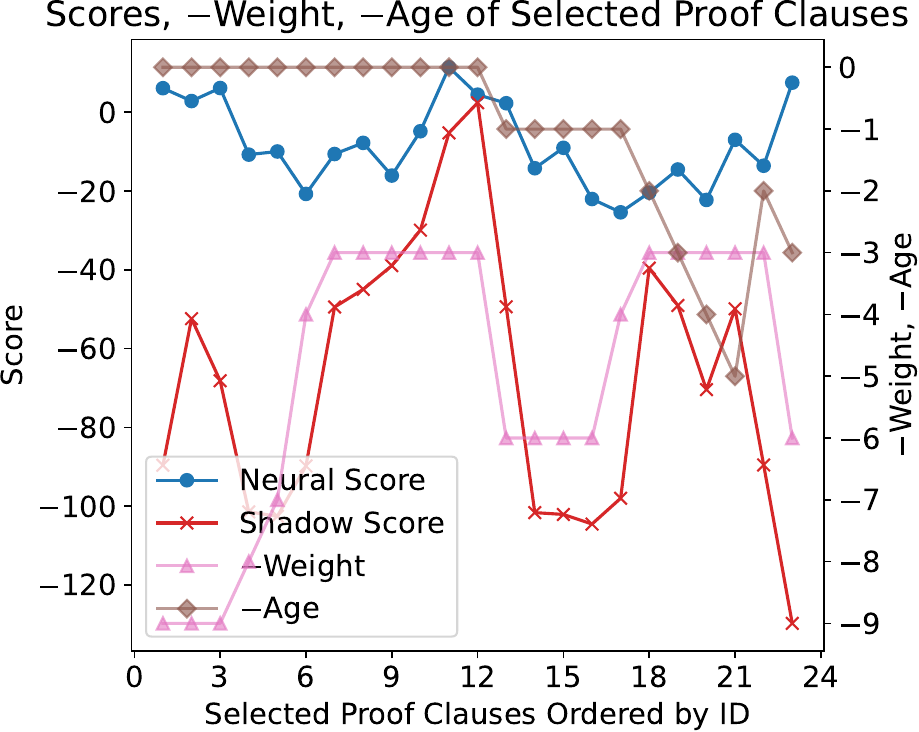}
  \caption{Scores of selected proof clauses from a run on \probtptp{CSR217+1}:
    The neural scores (y-axis, left) were computed by the best \tptp{}
    and the shadow scores by the best \coqhammer{}
    model, respectively. Standard clause weight and age are also shown (y-axis, right) for comparison;
    they are actually negated to align to with the  ``large means good'' polarity of the neural scores. }
  \label{fig:shadow:scores:csr217}
\end{figure}

\begin{figure}
  \centering
  \includegraphics[scale=0.5]{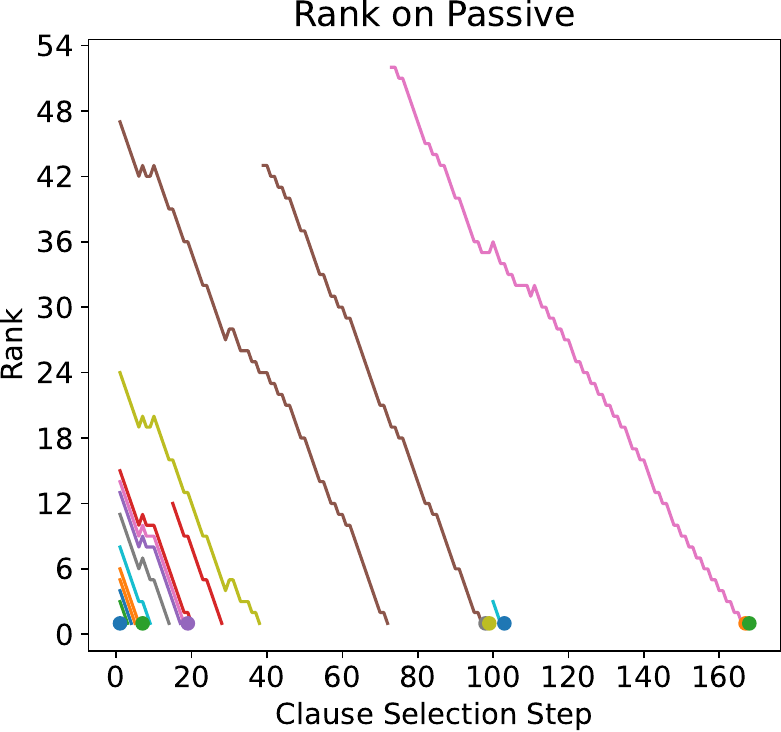}
  \includegraphics[scale=0.5]{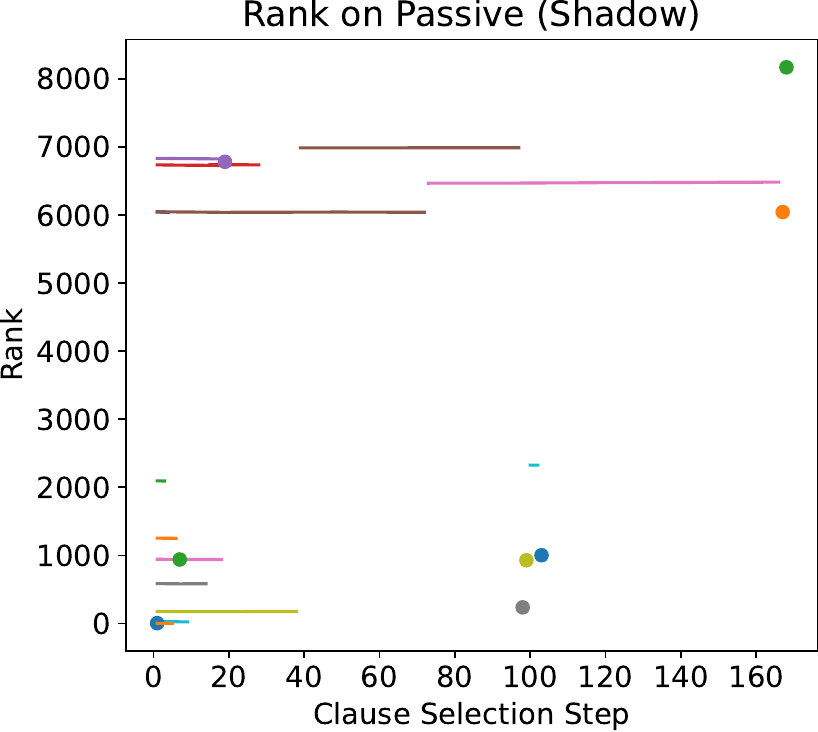}
  \caption{The same run on \probtptp{CSR217+1} is in Fig.~\ref{fig:shadow:scores:csr217}. 
   Left: The ranks of proof clauses on the passive set at each clause selection step. (A clause gets selected when it reaches rank 0).
   Right: The same passive set evolution, but the ranks are
    recomputed using the shadow model's scores. (Many proof clauses would not have a chance to get selected in a comparably short run guided by the shadow model.) 
    Note that the corresponding clauses share the same color on both pictures, but
    the same color may be reused for multiple clauses.}
  \label{fig:shadow:ranks:csr217}
\end{figure}

Since models do not generalize across datasets, it is natural to
analyze the situation by comparing how the models trained on different
datasets behave on individual problems. For example, we may wonder
whether a model fails because it incorrectly evaluates a few necessary
proof clauses, or because it incorrectly evaluates many of
them. Although both cases are possible, we commonly see the latter.

To analyze this situation, we visualize a run on a problem
(\probtptp{CSR217+1} from \tptp{}) in Fig.~\ref{fig:shadow:scores:csr217}. The
picture shows how the proof clauses that have been selected from the
passive set are evaluated by the neural model that guides the proof
search by providing \emph{neural scores} (a higher score means a
better clause). Moreover, every clause is also scored by a shadow
model, so named because it does not influence the proof search in any
way; it only computes \emph{shadow scores} for all the clauses. For
comparison, the classical metrics, weight and age, are also shown.

In this particular example, the neural and shadow scores are computed
by the best models trained on \tptp{} and \coqhammer{}, respectively. We can
see that the shadow model's scores of the proof clauses are much lower
than the neural model's scores.

In general, the scores computed by different models are not directly
comparable; what really matters is the induced ordering of clauses on
the passive set, which is shown in
Fig.~\ref{fig:shadow:ranks:csr217}. For each clause selection step,
the ranks (positions) of proof clauses on the passive set are
shown. The clauses that are added to the top of the passive set (i.e., get rank 0), and
hence get immediately selected, are depicted by dots. The left picture in
Fig.~\ref{fig:shadow:ranks:csr217} shows that the neural scores
computed by the \tptp{} model, which guides the proof search, result
in roughly 160 clause selection steps, with the worst rank of a proof
clause approximately 50. On the other hand, the right picture in
Fig.~\ref{fig:shadow:ranks:csr217} shows the same situation, but the
ranks on the passive set are recomputed using the scores of the shadow
model (\coqhammer). It shows that some of the proof clauses would be
ranked very poorly if the proof search was guided by the shadow
model. For example, the last proof clause in
Fig.~\ref{fig:shadow:scores:csr217} is favorably evaluated by the
\tptp{} model and hence immediately selected (the rightmost green dot
in Fig.~\ref{fig:shadow:ranks:csr217}). The same clause is scored poorly
by the \coqhammer{} model, so its position is $\sim$8000 when it is
added to the passive set.

\subsection{Transfer Between Models Trained on Mizar}
\label{sec:transf-between-miz}

\begin{figure}
  \centering
  \includegraphics[scale=0.5]{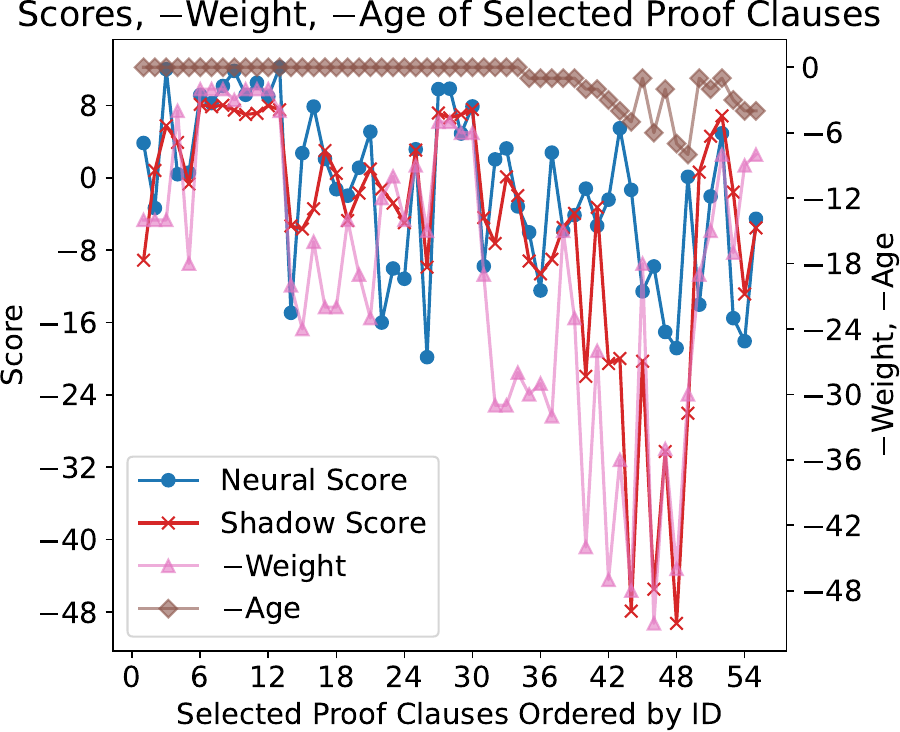}
  \caption{Scores of selected proof clauses from a run on \probtptp{SEU392+2}:
    The neural scores were computed by the best \tptp{}
    and the shadow scores by the best \mizar{}
    model.}
  \label{fig:shadow:scores:seu3922}
\end{figure}

\begin{figure}
  \centering
  \includegraphics[scale=0.5]{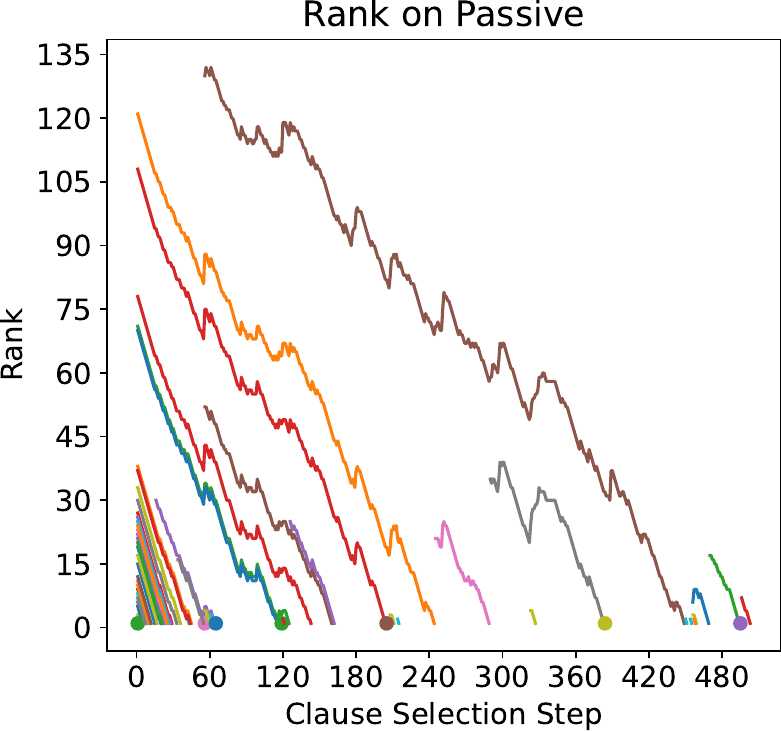}
  \includegraphics[scale=0.5]{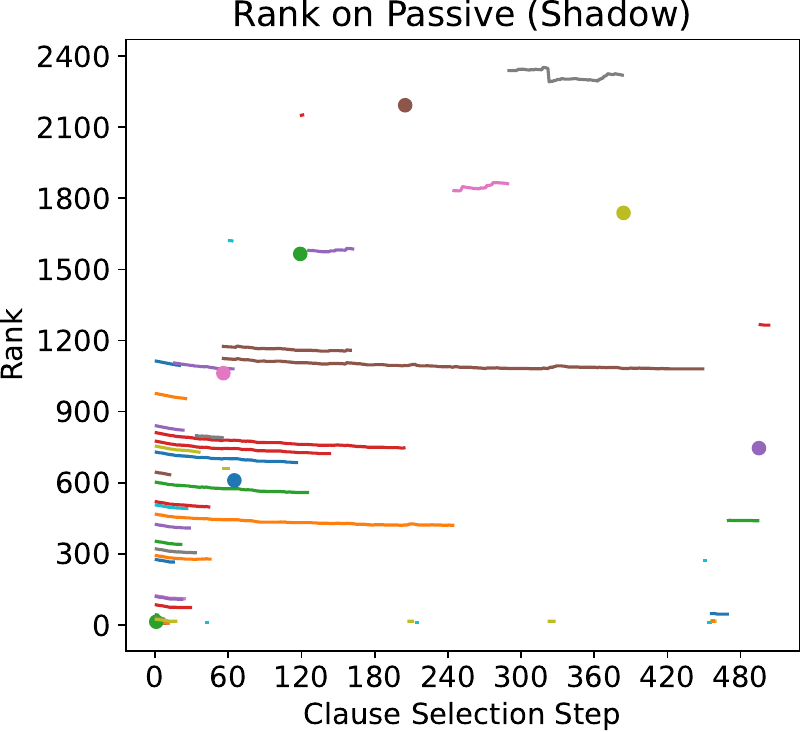}
  \caption{The same run on \probtptp{SEU392+2} as in Fig.~\ref{fig:shadow:scores:seu3922}: The neural scores and shadow
    scores are computed by the best \tptp{} model and the best \mizar{}
    model, respectively. The left picture shows the ranks of proof
    clauses on the passive set at each clause selection step. The
    right picture shows the same situation, but the ranks are
    recomputed using the shadow model's scores. Note that the
    corresponding clauses share the same color on both pictures, but
    the same color may be reused for multiple clauses.}
  \label{fig:shadow:ranks:seu3922}
\end{figure}

An interesting aspect of our datasets is that both \tptp{} and
\mizar{} contain problems exported from Mizar. For example,
\probtptp{SEU392+2}, from \tptp, contains a hammer version
(``chainy'') of a Mizar problem. In
Fig.~\ref{fig:shadow:scores:seu3922}, we see the evaluation of proof
clauses from the best \tptp{} (neural score) and \mizar{} (shadow
score) models; the proof search is guided by the \tptp{} model. Given
that the shadow model (\mizar) scores some heavy clauses poorly, it is
not surprising that Fig.~\ref{fig:shadow:ranks:seu3922} shows the
following: while the \tptp{} model (left picture) provides good
guidance, the ranks reevaluated with the \mizar{} model (right
picture) reveal that some clauses end up high on the passive
set. Accordingly, the proof attempt guided by the \mizar{} model
fails. A possible explanation is that the \mizar{} model is trained on
``bushy'' problems, which contain only the needed lemmata and theorems
used in the original human-supplied proof (see
Sect.~\ref{sec:datasets}).

\begin{figure}
  \centering
  \includegraphics[scale=0.46]{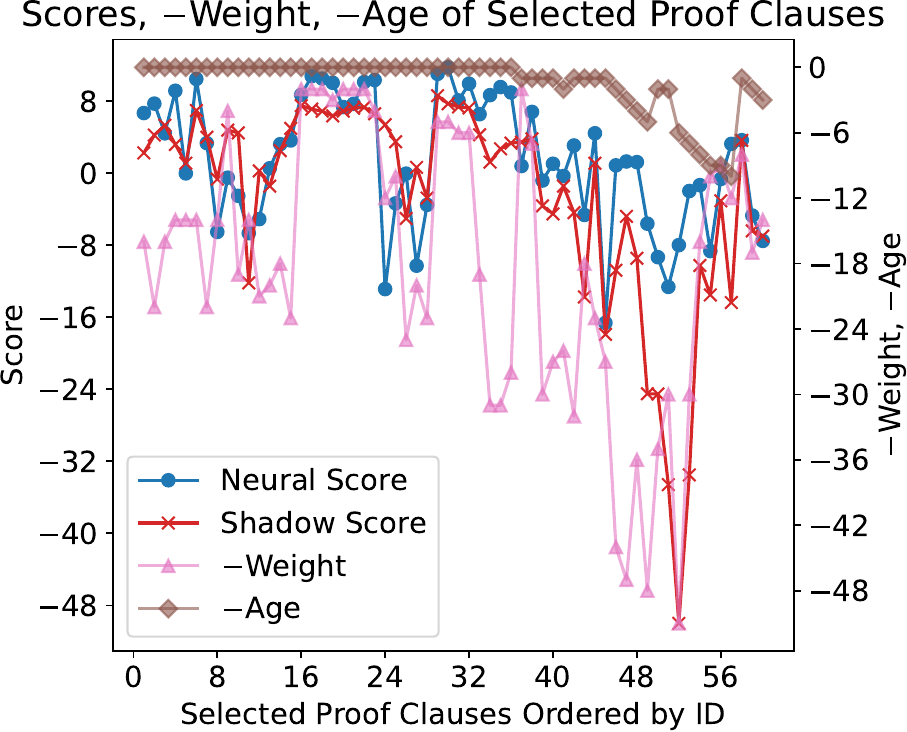}
  \includegraphics[scale=0.46]{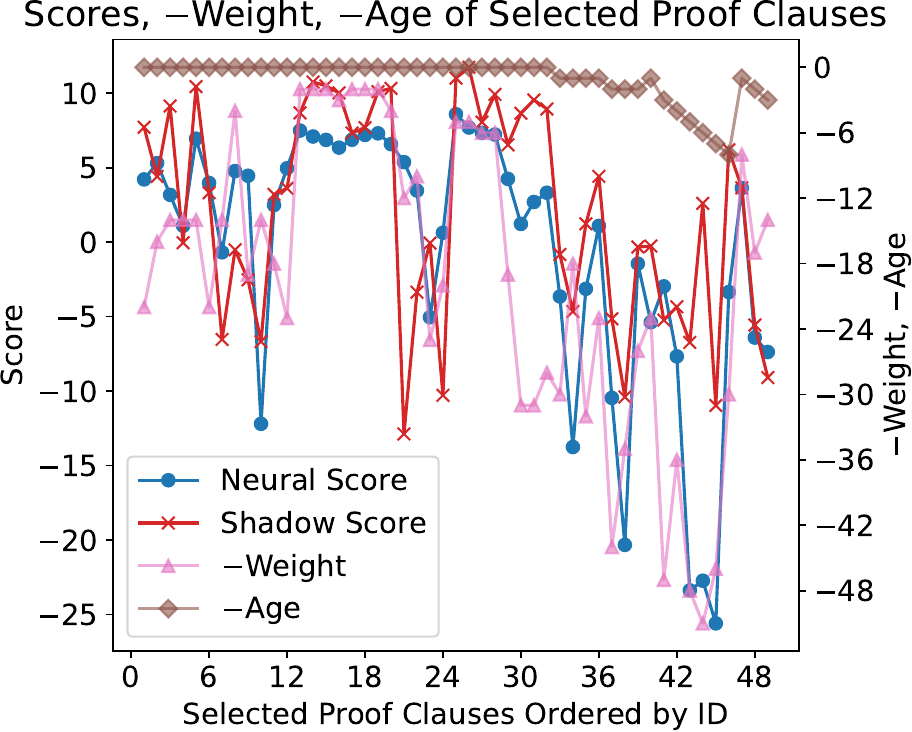}
  
  \caption{Scores of selected proof clauses from a run on
    \probtptp{SEU392+1}: In the left picture, the neural scores were
    computed by the best \tptp{} and the shadow scores by the best
    \mizar{} model. In the right picture, the models are swapped.}
  \label{fig:shadow:scores:seu3921}
\end{figure}

In fact, \tptp{} also contains the ``bushy'' version of the problem as
\probtptp{SEU392+1}. However, our \mizar{} training set does not
contain the corresponding problem.\footnote{Since both datasets use
  different MML versions, the \tptp{} problem \probtptp{SEU392+1}
  corresponds roughly to \probtptp{t12\_yellow19} on \mizar.} The left
picture in Fig.~\ref{fig:shadow:scores:seu3921} shows the same models
(\tptp{} guides and \mizar shadows) on this ``bushy'' problem. Again, the heaviest
clause, which has weight~51, is evaluated poorly by \mizar{}. However,
when the models are swapped, meaning \mizar{} guides the search and
\tptp{} shadows, a proof is also found, as the right picture in
Fig.~\ref{fig:shadow:scores:seu3921} shows. Clearly, the proofs differ
as the numbers of selected proof clauses show. Moreover, although in
both proofs there is a clause with weight~51, they are, in fact,
different clauses, and that explains the significantly different
scoring by the \mizar{} model (roughly $-50$ vs.~$-23$),
see~Fig.~\ref{fig:shadow:scores:seu3921}.

\begin{figure}
  \centering
  \includegraphics[scale=0.5]{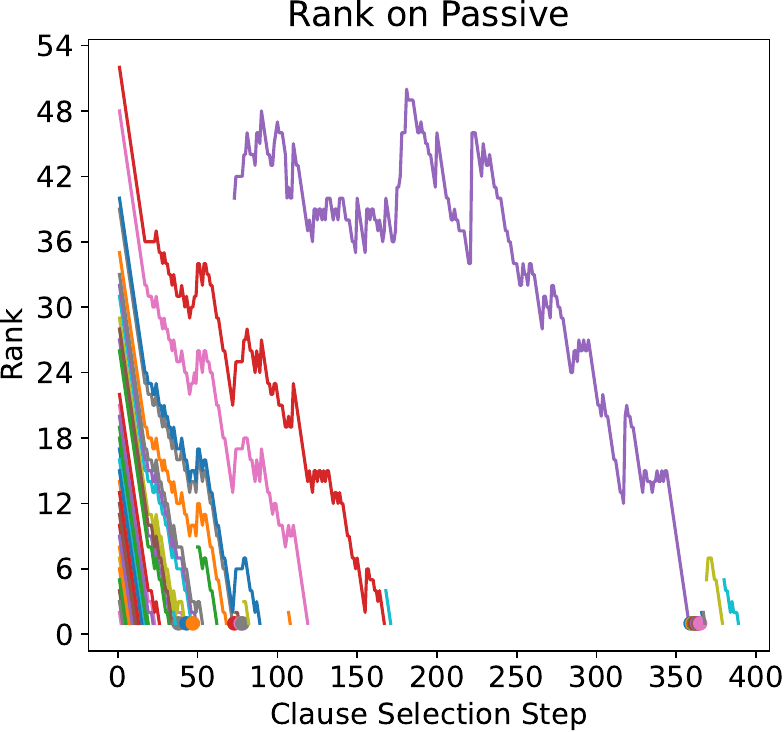}
  \includegraphics[scale=0.5]{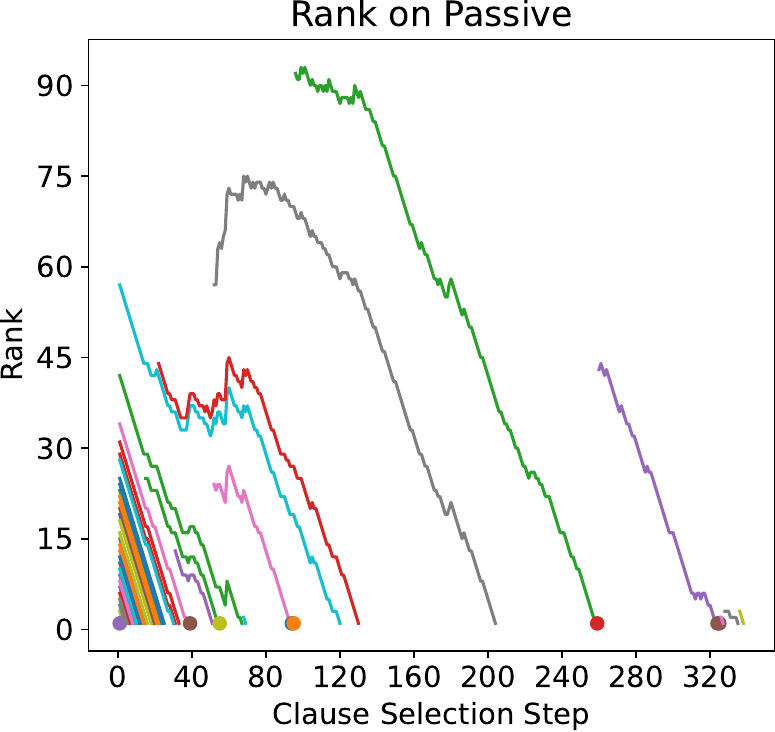}
  \caption{Problem \probtptp{SEU392+1}: The pictures show the ranks of
    proof clauses on the passive set at each clause selection step.
    In the left picture, the neural scores were computed by the best
    \tptp{} and the shadow scores by the best \mizar{} model. In the
    right picture, the models are swapped.}
  \label{fig:shadow:ranks:seu3921}
\end{figure}

Interestingly, although some clauses are scored poorly in the proof
guided by the \mizar{} model, this one requires only 338 activations
whereas the left one requires 389,
see~Fig.~\ref{fig:shadow:ranks:seu3921}. This picture also shows that
even a relatively poor score may result in a decent rank on the
passive set. We remark that when we use the best model trained on
\dataset{TPTP-noMizar}, it is still able to solve the problem, but it
requires 2034 activations.

\end{document}